\documentclass[a4paper,twocolumn,english,prl,showpacs,nofootinbib,preprintnumbers,amsmath,amssymb,superscriptaddress]{revtex4}
\usepackage[T1]{fontenc}
\usepackage[latin1]{inputenc}
\usepackage{graphicx}
\usepackage{amssymb}

\makeatletter

\usepackage{graphicx}
\usepackage{amssymb}

\usepackage{graphicx}  
\usepackage{dcolumn}   
\usepackage{bm}        
\usepackage[normalem]{ulem}
\usepackage{ae,aecompl}
\def\kbar{{\mathchar'26\mkern-9mu k}}

\usepackage{babel}

\usepackage{babel}
\makeatother
\begin{document}


\title{Reversible Destruction of Dynamical Localization}

\author{Hans Lignier}

\author{Julien Chab\'{e}}

\affiliation{Laboratoire de Physique des Lasers, Atomes et Mol\'{e}cules, UMR
CNRS 8523, Centre d'{\'E}tudes et de Recherches Laser et Applications,
Universit{\'e} des Sciences et Technologies de Lille, F-59655 Villeneuve
d'Ascq Cedex, France}

\author{Dominique Delande}

\affiliation{Laboratoire Kastler-Brossel, UMR CNRS 8552, Case 74, Universit{\'e} Pierre
et Marie Curie, 4 Place Jussieu, F-75252 Paris Cedex 05, France}

\author{Jean Claude Garreau}

\author{Pascal Szriftgiser}

\affiliation{Laboratoire de Physique des Lasers, Atomes et Mol\'{e}cules, UMR
CNRS 8523, Centre d'{\'E}tudes et de Recherches Laser et Applications,
Universit{\'e} des Sciences et Technologies de Lille, F-59655 Villeneuve
d'Ascq Cedex, France}

\date{\today}

\begin{abstract}
Dynamical localization is a localization phenomenon taking place,
for example, in the quantum periodically-driven kicked rotor. It is
due to subtle quantum destructive interferences and is thus of intrinsic
quantum origin. It has been shown that deviation from strict periodicity
in the driving rapidly destroys dynamical localization. We report
experimental results showing that this destruction is partially reversible
when the deterministic perturbation that destroyed it is slowly reversed.
We also provide an explanation for the partial character of the reversibility. 
\end{abstract}


\maketitle

Dynamical localization (DL) is one of the most dramatic manifestations
of how the quantum behavior of a complex system may differ from that
of its classical counterpart. It takes place in one-dimensional time-periodic
Hamiltonian systems where the deterministic motion is classically
chaotic and, on the average, equivalent to a diffusive expansion in
momentum space (the so-called chaotic diffusion behavior). Because
of subtle destructive interference effects, the quantum dynamics is
substantially different: while this dynamics is similar to the classical
one for short times, the diffusive behavior stops after some break-time
and the quantum momentum distribution gets frozen to a steady state
at long times. Interest in the DL come also from the fact that it
can be easily observed experimentally, e.g. by placing laser-cooled
atoms in a periodically kicked laser standing wave, the so-called
``kicked rotor''~\cite{Raizen:LDynFirst:PRL94}. The quantum
inhibition of classical transport is a rather generic behavior in
one-dimensional time-periodic Hamiltonian systems. It relies on the
existence of a class of states which are stationary under the one-cycle
evolution operator, the so-called Floquet states, forming a basis
of the Hilbert space. DL is thus a rather robust feature, which can
be observed for a large class of initial states, either pure states
or statistical mixtures. 

Another fascinating feature of DL is its sensitivity to external non-periodic
perturbations or deviations from the temporal periodicity of the system
\cite{Casati:BrisureLD:1991}. Various ways of breaking DL have been
studied experimentally and theoretically. One way is to add amplitude
noise to the kicks \cite{Raizen:LDynNoise:PRL98}. In such case,
it has been observed that DL is destroyed, i.e., that the quantum
motion remains diffusive at long times, as the classical motion. This
destruction has also been observed by adding a second series of kicks
at an incommensurate frequency \cite{AP:Bicolor:PRL00}, an experiment
that has also evidenced a very high sensitivity to frequency differences,
allowing observation of sub-Fourier resonances \cite{AP:SubFourier:PRL02}.
Another qualitatively different way of destroying DL is to introduce
a small amount of spontaneous emission in the system, thus breaking
its quantum coherences \cite{Raizen:LDynNoise:PRL98,Christ:LDynNoise:PRL98}.
While the first two examples correspond to a purely Hamiltonian evolution,
the latter one introduces an irreversible dissipative evolution.

In the case of a purely Hamiltonian dynamics, a fundamental question
remains, concerning the nature of the DL destruction: is this destruction
complete and irreversible or is it possible to stop the diffusive
behavior? Even better, is it possible to reverse the evolution and
reconstruct a more localized state? The purpose of this paper is to
report experimental results showing that such a relocalization is
possible (at least partially) when the non-periodic perturbation that
destroys DL is slowly reversed in time.

Let us first consider the standard kicked rotor Hamiltonian of a single
atom in a pulsed standing wave (SW): 
\begin{equation}
H_{0}=\frac{P^{2}}{2}+K\sin\theta\sum_{n=0}^{N-1}\delta_{\tau}(t-n),
\label{eq:H0}
\end{equation}
 where $P$ is the reduced momentum along the SW axis in units of
$M/(2k_{L}T_{1})$ ($k_{L}$ is the laser wavenumber and $M$ the
mass of the atom), $\theta=2k_{L}z$ the reduced position of the atom
along the SW axis, $K=\Omega^{2}T_{1}\tau\hbar k_{L}^{2}/(2M\Delta_{L})$
the kick strength ($\Omega$ is the resonant Rabi frequency of the
SW, $\Delta_{L}$ its detuning from the atomic resonance). The time
$t$ is measured in units of the period $T_{1}$ of the kicks. $N$
is the number of kicks, and $\delta_{\tau}$ is a Dirac-like function;
$\tau$ is the finite duration of the kicks. In the limit $\tau\rightarrow0$,
the dynamics of this Hamiltonian system is well known and depends
on only two parameters: $K$ and the effective Planck constant $\kbar=4\hbar k_{L}^{2}T_{1}/M$.
For $K\gg1$, the classical dynamics is a chaotic diffusion; a localized
set of initial conditions will spread in momentum space like a Gaussian
of width $\propto t^{1/2}.$ Below the break-time, the classical and
the quantum dynamics of an initially localized state are identical.
After the localization time, the quantum dynamics is frozen, the average
kinetic energy ceases to grow; at the same time, the momentum distribution
evolves from a characteristic Gaussian shape in the diffusive regime
to an exponential shape $\sim\exp(-|P|/L)$ (with $L$ being the localization
length) characteristic of the localized regime 
\cite{Shepelyanky:Kq:PD87,Cohen:LocDynTheo:PRA91}.

\begin{figure}
\begin{center}\includegraphics[%
  width=8cm]{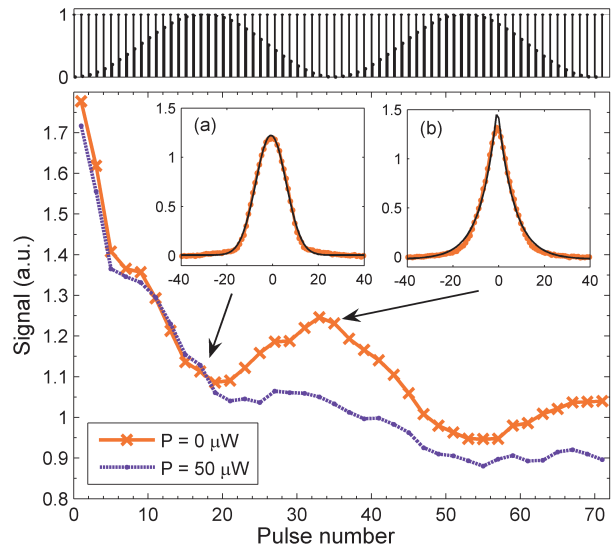}\end{center}

\caption{(color online) Upper frame: kick sequence. Main frame: Population
in the zero-velocity class as a function of the duration $N$ of the
pulse sequence with no resonant light (red crossed solid line) and
with a $50\mathrm{\mu W}$ pulse of resonant light applied at $t=17$
(blue dashed line). The absence of revival in presence of resonant light
(decoherence) is a clear-cut proof of the importance of quantum interference
for the reversibility of the DL destruction. The resonant pulse is
weak enough to induce no heating of the atoms but kills the phase
coherence, thus preventing the relocalization when the perturbation
reaches zero amplitude at $t=35$ and $t=70$. The inset (a) shows
the Gaussian velocity distribution at $t=17$, near the maximum of
the perturbation, in the absence of resonant light, inset (b) shows
a exponential shape near the minimum of the perturbation, $t=35$.}

\label{Fig:Revers2D}
\end{figure}

Consider now an experiment in which a slowly increasing and
then decreasing perturbation is added. This perturbation is added
to Hamiltonian (\ref{eq:H0}) as a second series of kicks with the
same frequency but with a time dependent amplitude (see upper frame
in figure~\ref{Fig:Revers2D}):

\begin{eqnarray}
H & = & H_{0}+\frac{K}{2}\sin\theta\left[1-\varepsilon\cos\left(\frac{2\pi t}{\Theta}\right)\right]\nonumber \\
 &  & \times\sum_{n=0}^{N-1}\delta_{\tau}\left(t-n-\frac{\phi}{2\pi}\right)
\label{eq:H}
\end{eqnarray}
where $\Theta\gg1$ is the period of the perturbation, $\phi$ the
relative phase between the two kicks series, and $\varepsilon$ the
modulation amplitude, with $\varepsilon\sim1$. Experimental values
are: $\Theta=35$, $\phi/2\pi=1/6$, and $\varepsilon=0.94$. At time
$t=N,$ the system has been exposed to $N$ kicks of the primary sequence
(with fixed strength $K$) and \textbf{$N$} kicks of the secondary
sequence (with time-varying strength), i.e., to a total of $2N$ pulses.

In order to experimentally realize the Hamiltonian Eq.~(\ref{eq:H}),
a sample of cold cesium atoms is produced in a standard magneto-optical
trap and released in the $F_{g}=4$ hyperfine ground-state sublevel.
A double sequence of $N$ pulses built according to Eq.~(\ref{eq:H})
is applied. The SW is detuned by $\Delta_{L}/2\pi=20$ GHz ($\sim3800\Gamma$,
where $\Gamma$ is the natural width of the atomic transition) with
respect to the $6S_{1/2},F_{g}=4\rightarrow6P_{3/2},F_{e}=5$ hyperfine
transition of the Cesium \emph{D}2 line ($\lambda_{L}=852$ nm). Such
largely detuned radiation essentially induces \emph{stimulated} transitions
responsible for conservative momentum exchanges with the atoms, so
that the dynamics is Hamiltonian. However, the SW laser line presents
a very broad low-level background (several hundreds of GHz) responsible
for a significant rate of dissipative spontaneous transitions. To get
rid of this problem, the SW passes through a 10 cm cesium cell before
interacting with the cold atoms. This filtering reduces the background
by more than one order of magnitude in a bandwidth of about 500 MHz
around the cesium transitions. Finally, after being transported by
a polarization-maintaining fiber, 92 mW of laser light, collimated
to a 1.5 mm waist, is available for the experiment, and retro-reflected
to build the SW. The frequency of each kick series is set to 30 kHz,
and the duration of each kick to $\tau=0.6\mathrm{\mu s}$. For these
values, the parameter $K$ is $\sim9$, and the localization time
$\sim10$ periods. The spontaneous emission rate is estimated to 0.06
per atom for the maximum duration of the experiment. Once the SW sequence
is over, the atomic momentum distribution is probed with a \emph{velocity
selective} Raman pulse. Thanks to Doppler effect, and a well-chosen
detuning, this pulse transfers the atoms in a well defined velocity
class from the hyperfine sublevel $F_{g}=4$ to the $F_{g}=3$ 
sublevel~\cite{Chu_RamanVSel:PRL91,AP_RamanSpectro_PRA02}.
The atoms remaining in the $F_{g}=4$ sublevel are pushed away by
a resonant laser beam. A resonant pulse brings the selected atoms
back in the $F_{g}=4$ level where their number is measured by a resonant
probe. The whole cycle starts again to measure the population in another
velocity class, allowing to reconstruct the full atomic momentum distribution.
Such a measurement is then performed for increasing pulse numbers
$N$.

A last precaution must however be taken. As discussed above, the SW
is intense enough to induce -- for a few atoms -- a real transition
from the level $F_{g}=4$ to the excited state, followed by spontaneous
emission leading possibly to the hyperfine level $F_{g}=3$, whatever
their momentum. Those atoms would be repomped to the $F_{g}=4$
sublevel and detected, generating an incoherent, $N$ dependant, background.
For each experiment, the Raman detuning is set very far away (at 10
MHz, more than one thousand recoil velocities), where the probability
to find a Raman resonant atom is very low. Except for this modification,
the experiment is launched in exactly the same conditions. The stray
background is corrected by subtracting the resulting signal from
the resonant one.

\begin{figure}
\begin{center}\includegraphics[%
  width=8cm]{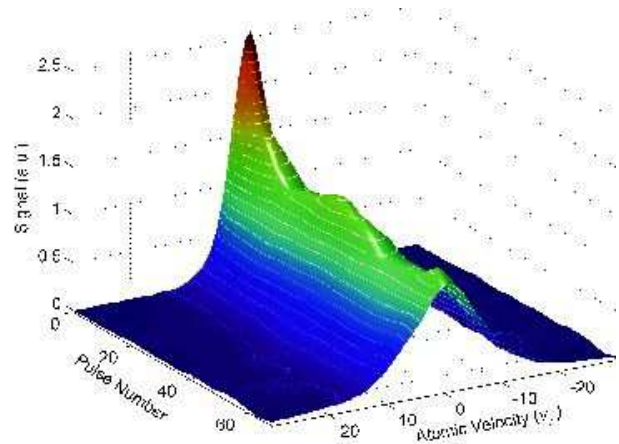}\end{center}

\caption{(color online). Experimentally measured velocity distributions as
a function of time. The atomic velocity is measured in units of the
recoil velocity $v_{r}=\hbar k_{L}/M\simeq$ 3.5 mm/s. At short times,
the diffusive broadening (or the reduction of the zero-velocity population)
of the velocity distribution is observed. When the slowly changing
perturbation is reversed (around $t=17$), the velocity distribution
\emph{starts to shrink.} This is highly non-trivial behavior,
showing that the destruction of dynamical localization by a slowly
time-varying kick sequence is reversible. After a second cycle of
the perturbation a second relocalization of the velocity distribution
is observed (around $t=70$). }

\label{Fig:Revers3D}
\end{figure}

Figure \ref{Fig:Revers3D} shows the velocity distribution as a function
of $N$. As expected, the early dynamics is diffusive. DL is expected
around $t=10$. Since the perturbation starts increasing from $t=0$,
DL is not visible and one could assume it is destroyed before it could
be seen. However, when the perturbation is reversed, the distribution
\emph{shrinks} and takes an exponential shape (see Fig.~\ref{Fig:Revers2D}),
signing a partial ``revival'' of the localization. This is clearly
visible in figure \ref{Fig:Revers2D} which displays the zero velocity
population $\Pi_{0}$ as a function of $N$ (red crossed solid line),
which is inversely proportional to the width of the distribution and
therefore directly proportional to the degree of localization. The
insets (a) and (b) display, respectively, the velocity distribution
at $t=17\sim\Theta/2$, where the perturbation reaches its maximum
amplitude, and at $t=\Theta=35$, where it is back to its minimum
initial value. In inset (a), the distribution is very well fitted
by a Gaussian, whereas the distribution in inset (b) is better fitted
by an exponential. The exponential shape at $t=\Theta$ is not the
only remarkable fact. The fact that the momentum distribution gets
narrower ($\Pi_{0}$ increases) is highly significative. Indeed, the
classical dynamics is diffusive and irreversible, forbidding, in the
general case %
\footnote{One can indeed conceive carefully prepared initial states that would
evolve to a narrower shape, but this is clearly not the case here.%
}, a return to a narrower distribution. Furthermore, DL leads to the
suppression of the classical diffusion, i.e., a freezing of the velocity
distribution; but it cannot lead to a narrowing of the distribution,
which is precisely what is experimentally observed in figure \ref{Fig:Revers2D}.
This is a key point of the present experiment: it proves that the
exponential shape observed at $t=\Theta$, where the perturbation
is zero, does not simply results from the DL that would be observed
in the periodic case, with no perturbation.

It is even possible to go further by proving the coherent nature of
the reversibility process. We have performed an additional experiment
where a resonant laser pulse irradiates the atomic cloud at $t=17$.
The intensity and detuning of this pulse are set such that, in the
average, only one or two spontaneous photons are emitted per atom.
Its purpose is to destroy the quantum coherences, with minimum of
heating and mechanical effects. The two curves in figure \ref{Fig:Revers2D}
(with and without the resonant pulse) are practically identical before
$t=17$, which indicates that heating resulting from the resonant
pulse is negligible %
\footnote{Each point in Fig.~\ref{Fig:Revers2D} represents a different experiment
with a different $N$. The resonant pulse of light is always applied at $t=17$
and the Raman detection then performed. Any significant heating effect
would enlarge the momentum distribution and would be detected.%
}. In the presence of resonant light, the ``revival\char`\" at
$t=35$ disappears almost completely (blue dashed line curve in figure
\ref{Fig:Revers2D}). Moreover, the velocity distribution (not shown
in Fig.~\ref{Fig:Revers2D}) is Gaussian around $t=35$. This clearly
proves that, in the absence of spontaneous emission, although DL is
not observed before $t=17$, there is a memory in the system which
is destroyed by spontaneous emission. This reinforces the idea that,
when the perturbation is reversed, a dynamically localized state is
recovered, at least partially. In fact, the ``revival'' of DL
is only partial, and a part of it is irremediably destroyed. As shown
in Figs.\ref{Fig:Revers3D} and \ref{Fig:Revers2D}, a second perturbation
cycle from $t=\Theta$ to $2\Theta$ has been performed, and a second
revival is observed. However, its amplitude is smaller than the first
one, and the shape of the velocity distribution is also damaged. 
This is due to fundamental reasons, although spontaneous
emission or experimental imperfections could also contribute to that.

A detailed discussion of the physical processes at work in our experiment
is beyond the scope of the present paper, and will be published elsewhere.
We give here a few guidelines to the theoretical interpretation. The
robust structure behind DL is the existence of Floquet states for
a time-periodic Hamiltonian system, which are eigenstates of the evolution
operator over one period. By their definition, such states repeat
identically (except for a phase factor) at each kick and thus do not
spread in momentum space. Any initial state can be expanded on the
complete set of Floquet states. Chaotic diffusion is -- in this
picture -- due to a gradual dephasing of the various Floquet states
(of different eigenenergies) that contribute to the initial state.
However, a non-trivial property of the periodically kicked rotor is
that all Floquet states are \emph{localized} in momentum space \cite{Casati:LocFloquetQKR:PRL90}:
this is the temporal analogous of Anderson localization in disordered
one-dimensional systems, as put on firm grounds in 
Ref.~\cite{Fishman:LocDynAnderson:PRA84}.
Only Floquet states localized close to the initial (zero) momentum
significantly overlap with the initial state and contribute to the
long term dynamics. At sufficiently long times, the various Floquet
states significantly contributing to the dynamics are completely dephased
(in a characteristic time which is but the break\textbf{-}time), the
momentum distribution covers all significantly populated Floquet states,
but cannot extend further in momentum space, leading to the freezing
of the diffusive growth. When the dynamics is no longer exactly periodic,
population is transferred among the various Floquet states, and Floquet
states localized farther from $P=0$ can be populated. In this situation,
DL is thus expected to be destroyed. There is however a situation
where such an evolution can be controlled: if, at any time, the Hamiltonian
is almost periodic with, for example, a kick strength $K(t)$ slowly
changing with time $t,$ an adiabatic approximation can be used. The
atomic state at time $t$ can be expanded in terms of the ``instantaneous\char`\"
Floquet eigenbasis corresponding to the local value of $K(t)$. If
the variation of $K(t)$ is slow enough, the evolution is adiabatic
in the Floquet basis, meaning that the populations of the Floquet
eigenstates do not change with time, while the eigenstates themselves
evolve \cite{Delande:FloquetAdiab:JPB95,Hone:FloquetAd:PRA97}. This
leads to an apparent diffusive broadening of the momentum distribution
\cite{AP:SubFMecs:EPL05}, but the robust Floquet structure is still
underlying. To recover the localization, it is sufficient to reverse
the evolution of $K(t)$ back to its initial value. One then recovers
the initial well localized Floquet eigenstates with unchanged populations,
i.e., a dynamically localized momentum distribution. This is the deep
origin of the revival of the localization experimentally observed
above. Any phenomenon breaking phase coherence (such as a spontaneous
emission) will redistribute the atomic wavefunction over other Floquet
states, eliminating all possibility of revival. However, the revival
is only partial, because the evolution cannot be made 100\% adiabatic.
Indeed, even for very slow changes of $K(t),$ there are some avoided
crossings between various Floquet states of such size that they will
be crossed neither diabatically, nor adiabatically, and will consequently
redistribute the population over the Floquet states, partially destroying
the reversibility.

To summarize, we have observed that the destruction of dynamical localization
in the kicked rotor, induced by a non-periodic driving can be partially
reversed. If the external driving evolves sufficiently slowly, some
information is carefully stored in the populations of the various
Floquet states. Although it is not visible in the momentum distribution
-- which seems to follow an irreversible diffusive broadening -- it
can be easily restored by reverting the driving back to its
initial value, producing a relocalization of the wavefunction.
We have also show that this intrinsically quantum behavior is destroyed
by decoherence, i.e., by adding spontaneous emission to the experiment.

\begin{acknowledgments}
Laboratoire de Physique des Lasers, Atomes et Mol\'{e}cules (PhLAM)
is Unit\'{e} Mixte de Recherche UMR 8523 du CNRS et de l'Universit\'{e}
des Sciences et Technologies de Lille. Centre d'Etudes et de Recherches
Laser et Applications (CERLA) is supported by Minist\`{e}re de la
Recherche, R\'{e}gion Nord-Pas de Calais and Fonds Europ\'{e}en
de D\'{e}veloppement {\'E}conomique des R\'{e}gions (FEDER). Laboratoire
Kastler-Brossel de l'Universit\'{e} Pierre et Marie Curie et de l'{\'E}cole
Normale Sup\'{e}rieure is UMR 8552 du CNRS. CPU time on various computers
has been provided by IDRIS. 

\end{acknowledgments}


\end{document}